\documentclass{article}[16pt]
\usepackage[utf8]{inputenc}
\usepackage{graphicx}
\begin{document}
\begin{center}

{\Large LARGE DIFFUSE DWARFS IN THE DYNAMICALLY COLD TRIPLE GALAXY SYSTEMS} \\

 \bigskip
 
 {\large I.D.KARACHENTSEV$^{1}$, 
A.E.NAZAROVA$^{1}$,
V.E.KARACHENTSEVA$^{2}$} 

$^{1}$Special Astrophysical Observatory of the Russian Academy of Sciences, N.Arkhyz, KChR, 369167, Russia, idkarach@gmail.com

$^{2}$Main Astronomical Observatory, National Academy of Sciences of Ukraine, Kiev, 03143, Ukraine, 
valkarach@gmail.com

\bigskip

{\bf Abstract}
 \end{center}

We report on the discovery of three large diffuse dwarf (LDD) galaxies located in isolated triple systems.
They have effective diameters of (3.6--10.0) kpc and effective surface brightness of 
(26.2--27.3)$^m/$sq.arcsec. We note that the LDD galaxies tend to occur in small groups with a very low
dispersion of radial velocities. The total (orbital) mass of the triplets approximately equals to their 
integral stellar mass within velocity measurement errors. The presence of LDD galaxies in cold multiple
systems seems mysterious.

{\bf Key-words}: galaxies~--- dwarf galaxies~--- low surface brightness galaxies 
   
\section{Introduction}
Over a wide range of luminosities, the average surface brightness of galaxies, $SB$, and their
integral absolute magnitude, $M$, follow a relation
$SB = (1/3) M +$ const, which corresponds to an approximate constancy of the average
volumetric stellar density for major and dwarf galaxies [1].
However, with
the advent of deep sky surveys, a specific population of low
surface brightness galaxies has been discovered,
whose luminosity is typical of dwarf systems and whose sizes are comparable to those of
normal galaxies. These objects are called ``ultra-diffuse
galaxies'' (UDG). As defined by van Dokkum et al. [2], 
these include galaxies with a central
surface brightness in the $g$-band $SB_g(0)>24$~$^m/$sq.arcsec
and a linear effective diameter $A_{50} > 3.0$~kpc, within which half of
the galaxy’s luminosity is contained. 

Many UDG galaxies have been discovered in the nearby clusters: Virgo [3], 
%(Mihos et al. 2017), 
Fornax [4] %(Munoz et al. 2015) 
and Coma [5], %(Koda et al. 2015), 
and a small number have been also found in nearby groups around NGC\,253, Cen\,A, NGC\,5485 [6,7,8]. 
A catalog of 7070 UDG candidates selected over 20\,000~sq.degr. of sky was recently published by Zaritsky et al. [9].%\citet{zar2023}.
%(Crnojevic et al. 2016, Toloba et al. 2016, Merritt et al. 2016). 
According to [10], %Roman & Trujillo (2017), 
about 40\% of UDG objects are found
in clusters, about 20\% are located in groups, and
the remaining 40\% occur in scattered filaments, avoiding common field regions. No isolated
UDG galaxies have yet been discovered. This
arrangement of diffuse galaxies relative to the elements of the cosmic web indicates that the
structure of UDG galaxies is determined not so much
by features of their internal evolution as by the influence of external environment. The disperse
stellar structure of UDG galaxies is obviously a
sensitive indicator of the tidal influence of their neighbors.

Studying the closest examples of UDG galaxies helps to advance our understanding of their
specifics. In the Local Volume with a radius of 10~Mpc,
15 objects were noted [11] %(Karachentsev et al.2017) 
that meet the criterion of an ultra-diffuse galaxy. 
All of these UDG objects are located in the
known nearby groups. Among them, some (Sag\,dSph, KK\,208, Scl-MM-Dw2, And\,XIX) have
an elongated structure with an apparent axial ratio
$b/a < 0.5$, which is caused by gravitational disturbance from a massive neighbor. Others
(Garland, NGC\,3521sat, d0226+3325) have an irregular
shape and a young stellar population, indicating that these dwarfs likely formed from tidal tails
and bridges on the outskirts of major galaxies.
We consider it appropriate to strengthen the criterion for a large diffuse dwarf (LDD) galaxy,
selecting into this category objects of even lower
surface brightness, having a round smooth shape and lack of young stellar population. As a
criterion for a galaxy to belong to a LDD object, we
use the following conditions:

i. The effective diameter of a galaxy in the $g$-band is $A_{50,g} > 3.0$ kpc;

ii. Effective surface brightness in the $g$-band $SB_{50,g} > 26.0^m$/sq.arcsec;

iii. Apparent axial ratio $b/a > 0.5$;

iv. Morphological type dSph with a smooth shape and an old population ($g - r > 0.50$).

Of the thousand galaxies in the Local Volume, only five known galaxies satisfy these
conditions: Cen-MM-dw1, IKN, KK\,77, Cen-MM-dw3 and
NGC\,4631dw1 with distances in the range (3.6--7.4)~Mpc, effective diameters $A_{50} = (3.1 -
5.6)$~kpc, effective surface brightness $SB_{50} =
(26.2 - 28.1)^m$ /sq.arcsec and absolute magnitudes $M_B = (-11.6^m - -12.6^m)$. The names
of the galaxies are indicated as they are presented
in the Updated Nearby Galaxy Catalog [1], %(Karachentsev et al. 2013), 
a regularly updated
version of which is available on-line\footnote{http://www.sao.ru/lv/lvgdb}. It is likely that the
Local Volume contains other LDD galaxies that have not yet been discovered due to the
incompleteness of deep sky surveys.
\begin{figure}
\includegraphics[height=6cm]{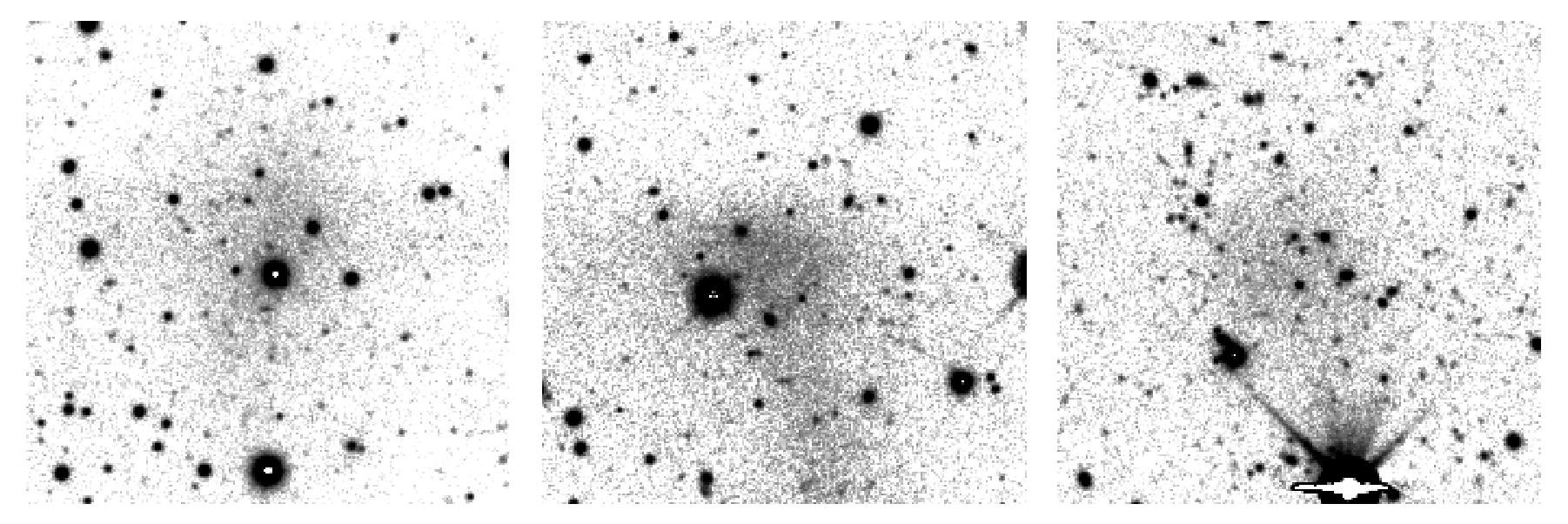}    
\caption{Images of three large diffuse dwarf galaxies from the DESI Legacy Imaging Surveys: 
LDD\,0954-28, LDD\,0911-14 and LDD\,0852-02 from the left ro right. Each image size is 
$2^{\prime}\times 2^{\prime}$. North is to the top, East is to the left.}  \label{figure1}
\end{figure}

\section{A LDD galaxy in the NGC\,3056 triplet}
While searching  for new nearby dwarf galaxies in DESI Legacy Imaging Surveys, DR10 [12], %Dey et al. 2019), 
we discovered a low surface brightness
object at coordinates 09:54:43.9 --28:30:54 (J2000). Its image is shown on the left panel of Fig.1.
To the north of it at a distance of $13^{\prime}$, there is a
galaxy of type S0a with a radial velocity $V_{LG} = 674$~km~s$^{-1}$ relative to the centroid of the Local
Group. This galaxy with its two satellites:
ESO\,435-016 and ESO\,435-020 forms a triple system, presented in the list of nearby isolated triplets
of galaxies [13]. %(Makarov & Karachentsev 2009). 
The parameters of this system are shown in Table 1. Its columns contain: galaxy name; its
coordinates; morphological type; apparent $B$-magnitude;
radial velocity in km~s$^{-1}$; projection separation from the main galaxy, $R_p$, in kpc; estimate of the
orbital (projected) mass $M_p=(16/\pi G)\cdot \Delta V^2\cdot R_p$
in units of $10^{10}M_{\odot}$, where $\Delta V$ is the difference in radial velocities of the satellite and
the host galaxy, and $G$ is the gravitational constant [14].
\begin{table}
\caption{Properties of  NGC\,3056 triplet of galaxies.}  \label{table1}
\begin{tabular}{lclclrr} \hline

  Name         &RA(2000.0)DEC,      &  Type     &  B,        &   $V_{LG}$,  & $R_p$,&   $M_p$, \\ 
               & deg\,\,\,\,\,deg   &           &   mag     & km s$^{-1}$ & kpc  & $10^{10}$ \\ \hline
 NGC 3056      & 148.637 -28.298   &    S0a    &  12.6     &  674$\pm$5    &    0 &       ---     \\
 ESO 435-016   & 149.691 -28.622   &    Im     &  13.5     &  678$\pm$4    & 209  &   0.4          \\
 ESO 435-020   & 149.838 -28.133   &    Irr-p  &  14.4     &  673$\pm$2    & 227  &    0.03        \\
 LDD 0954-28   & 148.683 -28.515   &    Sph    &  17.8     &    ---        &  47    &          ---\\
\hline
\end{tabular}
\end{table}
%(Karachentsev & Kashibadze 2021). 
The irregular galaxy ESO\,435-020 has a peculiar
structure with signs of recent merging. A feature of the triplet
is the small dispersion of the radial velocities of galaxies, comparable to the errors in velocity
measurements. The reason for this may be the
projection effect, when the velocity vectors of both satellites are almost perpendicular to the
line of sight. The observed low velocity dispersion may
also be a consequence of the low total mass of the galaxy triplet.

The last row of the table corresponds to the LDD galaxy we noted with a very low surface
brightness. We assume that this object is a physical
member of the triplet. The analysis undertaken by Karachentseva et al. [15], %Karachentseva et al. (2010), 
showed that
isolated dSph galaxies are extremely rare. In the
volume of the Local Supercluster with a radius of approximately 40~Mpc, only a dozen
such putative cases have been noted. Radial velocity
measurements in dSph galaxies are very difficult due to the absence of a noticeable amount of
neutral hydrogen in them and due to the low optical
surface brightness. In those rare cases when such measurements were possible, the radial
velocities of spheroidal dwarfs turned out to be close to
the velocities of neighboring normal galaxies [16], %(Sharina et al. 2016), 
making the assumption of their isolation unlikely.

\section{Other examples of LDD in triplets of galaxies}
The catalog of isolated galaxy triplets in the Local Supercluster [17] %(Makarov & Karachentsev2009) 
contains data on 168 triple systems with radial
velocities $V_{LG}<3500$~km~s$^{-1}$. The sample of these triplets is characterized by the following
median parameters: member radial velocity dispersion of
40~km~s$^{-1}$ , projected harmonic radius of 155 kpc, projection (orbital) mass of $5\cdot 10^{11} M_{\odot}$, 
and orbital mass-to-stellar mass ratio $M_p/M_* = 25$.

We assumed that a low velocity dispersion in a multiple galaxy system may be a favorable
factor for the presence of very diffuse objects in it. Among
168 nearby triplets, there are 17 systems with the small ratio $M_p/M_* < 2$, located in the Legacy Imaging Surveys area.
Looking at these cases, we found two more triple systems with major members NGC\,2781 and UGC\,4640, containing candidate LDD objects. Their
images are shown in the middle and right panels of Fig.1. Data on these triplets are presented
in Tables 2--3, the parameters of which are similar to
those in Table 1. Radial velocities of galaxies and their errors are taken from Lyon
Extragalactic Database = LEDA [18]. % (Makarov et al. 2014). 
The distances to NGC\,2781 (30.6~Mpc) and UGC\,4640 (49.0~Mpc) were estimated from their radial velocities
taking into account local cosmic flows in the Numerical
Action Method model [19]. %(Shaya ey al. 2017). 
The distance to NGC\,3056 (12.2~Mpc) was
determined from surface brightness fluctuations [20]. %(Tonry et al. 2001).
\begin{table}
\caption{Properties of  NGC\,2781 triplet of galaxies.}  \label{table2}
\begin{tabular}{lclclrr} \hline
 Name         &RA(2000.0)DEC,      &  Type     &  B,        &   $V_{LG}$,  & $R_p$,&   $M_p$, \\ 
              & deg\,\,\,\,\,deg   &           &   mag     & km s$^{-1}$ & kpc  & $10^{10}$ \\ \hline
 NGC 2781     &  137.864 -14.817  &    S0a    &  12.5    &1766$\pm$22   &    0  &     ---   \\
 DDO 57       &  137.832 -15.051  &    Im     &  14.8    & 1784$\pm$2   &    94 &   3.6     \\
MCG-02-24-03  &  138.028 -15.432  &    Sm     &  15.2    & 1794$\pm$6   &    101&    9.3    \\
LDD 0911-14   &  137.856 -14.703  &    Sph    &  18.2    &      ---       &    46 &     ---   \\
 \hline
 \end{tabular}
 \end{table}
 
 \begin{table}
\caption{Properties of UGC\,4640 triplet of galaxies.}  \label{table3}
\begin{tabular}{lclclrr} \hline
 Name         &RA(2000.0)DEC,      &  Type     &  B,        &   $V_{LG}$,  & $R_p$,&   $M_p$, \\ 
              & deg\,\,\,\,\,deg   &           &   mag     & km s$^{-1}$ & kpc  & $10^{10}$ \\ \hline
 UGC 4640     &   132.933 -02.134 &      Sc   &    13.8   &   3091$\pm$3  &     0&       ---\\
 Arp 257a     &   132.909 -02.367 &     Sm    &  14.4     & 3103$\pm$4    & 200  &  3.4 \\
 Arp 257b     &   132.908 -02.354 &     Im    &   16.8    &  3106$\pm$6   &  189 &   5.0 \\
 LDD 0852-02  &   133.148 -02.177 &     Sph   &  19.2     &     ---       &  184    &       --- \\
\hline
 \end{tabular}
 \end{table}

As a control sample, we searched for LDD galaxies in the virial zones of 17 triplets with an $M_p/M_* > 100$ and 
didnot find a single LDD object. Since triplets of galaxies with $M_p/M_* <2$ constitute only 10 percent of their
total number (17/168), the probability of three triplets with LDD members falling into this category is 0.001.

\section{Surface photometry of LDD galaxies}
We performed surface photometry of three new very low surface brightness galaxies, absent in [9] catalog, to
estimate their structural parameters. For this purpose, data from
DESI Legacy Imaging Surveys, DR10 in the $g$ and $r$ bands were used. The photometry of the galaxies
was carried out by measuring photometric curves of growth using standard ellipse-fitting and aperture photometry 
techniques in $photutils\footnote{https://photutils.readthedocs.io/en/stable/}$. This was preceded by background subtraction. 
Also, foreground and bright background objects were masked and then the corresponding pixels were replaced by the mean 
flux in the aperture rings contained in the mask. We fit the resulting curve of growth, $f(r$), using the following modified exponential law:

\begin{equation}
	f(r) = f(r_0) + f^{\mathrm{exp}}(r), \ \ r > r_0
\end{equation}
where $f^{\mathrm{exp}}(r)$ -- the flux corresponding to the standard exponential law and $f(r_0)$ -- the additional flux
from the inner part ($r < r_0$) of the galaxy. This leads to:

\begin{equation}
	f(r) = f_{\mathrm{tot}} \left(1 - \frac{f^{\mathrm{exp}}_{\mathrm{tot}}}{f_{\mathrm{tot}}}\left(1 + \frac{r}{h}\cdot\mathrm{e}^{-r/h}\right)\right)
\end{equation}
where $f_{\mathrm{tot}}, f^{\mathrm{exp}}_{\mathrm{tot}}$ and $h$ -- fitting parameters of the model: 
total flux, total flux (without adding $f(r_0)$) from the standard exponential law and the exponential scale from the standard exponential law.

The results of measuring the integral magnitudes of galaxies $g$ and $r$,
effective radii $r_{50,g}$ and $r_{50,r}$ are presented in Table~4. It
also shows $B=g+0.542\cdot(g-r)+0.141$ and $V=g-0.496\cdot(g-r)-0.015$ integral magnitudes of
galaxies in $B$ and $V$ system [21]. %(Abott et al. 2021).
\begin{table}
\caption{Photometric parameters of the LDD galaxies.} \label{table4}
\begin{tabular}{lccccccccccc} \hline  
Name        &  $g$,    &  $r$,   &  $B$,  &  $V$, &   $r_{50,g}$,     &$r_{50,r}$,       &$SB_{50,g}$,    & $A_{50,g}$, & $B-V$,  &  $E(B-V)$, &    $M_B$  \\ 
             &mag     &  mag   &  mag  &  mag  &$^{\prime\prime}$&$^{\prime\prime}$& mag/sq.arcsec&    kpc       & mag    &  mag       &    mag \\                
\hline
LDD0954-28  & 17.31   & 16.75  & 17.75  &17.02 &  30.77 &  31.96  &  26.74   &   3.64  &   0.73 &   0.065 &   --12.95 \\
LDD0911-14  & 17.76   & 17.12  & 18.25  &17.43 &  19.83 &  19.65  &  26.24   &   5.88  &   0.64 &   0.046 &   --14.37 \\
LDD0852-02  & 18.70   & 18.08  & 19.18  &18.38 &  21.10 &  21.15  &  27.31   &  10.02  &  0.62  &  0.016  &  --14.34  \\
\hline
 \end{tabular}
 \end{table}

As one can see, the integrated color indices $B - V$ , taking into account the color excess $E(B -
V)$ due to extinction, turns out to be typical for dSph galaxies with old stellar population.
The effective linear diameters $A_{50} > 3.0$~kpc of all three diffuse galaxies and their effective
surface brightness $SB_{50} > 26.0^m$/sq.arcsec satisfy the
conditions formulated above for LDD galaxies. The assumption that these LDD galaxies are
physical members of the triplets under consideration looks very plausible, given that such diffuse objects have not yet been discovered in the
general field. The absolute magnitudes of these LDD galaxies,
indicated in the last column of Table 4,  are 1--3~mag
brighter than similar objects in the Local Volume.

\section{Brief discussion}
We estimated the stellar mass of galaxy triplets $M_*$ using the ratio  $M_*/M_{\odot}=0.6\cdot(L_K/L_{\odot}$) 
according to Lelli et al. [22].  %Lelli et al. 2016. 
The values of the total
luminosity of triplets in the $K$-band are taken from the catalog [17]  %(Makarov & Karachentsev 2009)
with correction for the adjusted distance. Data on Table~5  show
that the estimate of the total (projected) mass of triplets turned out to be approximately equal
to their stellar mass within the $M_p$ errors due to the velocity
measurement errors. From this we can conclude that the ``cold'' kinematics of the triplets under
consideration does not require the presence of a noticeable
amount of dark matter in them. It remains unclear how such a feature of triplets can be related
to the presence in their volume of very diffuse galaxies with
old stellar populations. 
\begin{table}
\caption{Triple systems of galaxies with LDD.}  \label{table5}
\begin{tabular}{lccc}\hline
 Name     &    $D$,     &  $\log(M_*)$,   &   $\log(M_p)$, \\     
          &    Mpc   &   $10^{10}$    &    $10^{10}$    \\  
\hline
 NGC 3056 &   12.2   &    0.87    &      0.22$\pm$0.20 \\        
 NGC 2781 &  30.6    &   5.11     &     6.45$\pm$2.85    \\    
 UGC 4640 &   49.0   &    2.51    &      4.20$\pm$0.80     \\    
  \hline
   \end{tabular}
   \end{table}

In general, various possible scenarios for the formation of LDD galaxies have been discussed in
the literature: a high angular momentum of the LDD [22], %(Amorisco & Loeb, 2016), 
a stellar feedback from the
host galaxy [23], %(Di Cintio et al, 2017), 
and ``failed Milky Way'' mechanism [24]. %(van Dokkum et al, 2016).

Note that the average projection separation of LDD galaxies, 92~kpc, is
about half the average distance of late-type satellites, 170~kpc.
The same effect of segregation of dSph and dIrr galaxies is also well known in other groups
and clusters of galaxies.

The diffuse satellite LDD\,0911-14 exhibits a strong shape distortion in the form of a tidal tail
directed towards the massive host galaxy NGC\,2781. Our
photometry of this satellite was limited to the main body of the object. Taking into account the
tidal tail almost doubles the integral luminosity and effective
diameter of this galaxy. 

It is obvious that galaxy systems with cold kinematics and the
presence of very diffuse members can also be found among groups
with a larger population. As an example, we note a group of four satellites around the galaxy
NGC\,660. At a projection distance of 13$^{\prime}$ east of NGC\,660
there is an extremely low surface brightness galaxy (01:43:55.2+13:38:42), discovered by Karachentsev \& Kaisina [25]. 
%Karachentsev & Kaisina (2022 ). 
The spiral galaxy NGC\,660 itself
has a very peculiar shape in the form of two merging galaxies. With its low ratio $M_p/M_*\simeq 3$,
this group stands out among other groups in the Makarov \& Karachentsev [13] %Makarov & Karachentsev (2011) 
catalog.

Recently, Okamoto et al. [26] %Okamoto et al. (2024) 
discovered an extremely low surface brightness satellite near
the nearby spiral galaxy NGC\,253 using deep stellar photometry
with the Hyper Suprime-Cam on the Subaru telescope. According to the authors, this ``ghost''
galaxy has an effective diameter of $6.7\pm0.7$~kpc, an effective
surface brightness of $SB_{50}\sim 30^m/$ sq.arcsec, an apparent axial ratio of $b/a = 0.94$ , and
old stellar population. This satellite, NGC\,253-SNFC-dw1, is
located at a projection separation of $R_p = 75$~kpc from the center of NGC\,253 and shows
weak signs of tidal disruption. In its size and extremely low surface
brightness, this dim satellite of NGC\,253 is similar to the Milky Way satellite, Antlia\,2, with
$A_{50}=5.8\pm0.6$~kpc, $SB_{50} =31.9^m/$sq.arcsec [27]
%(Torrealba et al. 2019) 
and the satellite of M\,31, And\,XIX , with $A_{50} = 6.2\pm2.0$~kpc, $SB_{50} =
31.0^m/$sq.arcsec [28]. % (Martin et al. 2016). 
Such objects are undetectable by conventional photometry, since their surface brightness is 5--6 mag fainter
than that of the UDG galaxies discussed by [3]. %van Dokkum et al. (2015).

It is interesting to note that the galaxy NGC\,253, along with NGC\,2683 and NGC\,2903, has the
minimum radial velocity dispersion of satellites ($\sigma_v < 45$~km~s$^{-1}$)
among the 25 brightest galaxies in the Local Volume with a luminosity similar to that of the
Milky Way. This fits with the trend that the cold kinematics of the
satellites (or the deficiency of dark matter in the group) favors the survival in the group of
``fragile'' satellites with very low stellar density.

We did not consider here the possible reasons for the observed correlation between the
presence of LDD galaxies in a group and the deficiency of dark matter in it.
Apparently, it is necessary to accumulate richer statistics of such cases, as well as perform
dynamic modeling of tidal destruction of diffuse satellites under different
assumptions about the shape of the satellites’ orbits and the amount of dark matter in the LDD
galaxy. According to Penarrubia et al. [29], %Penarrubia et al. (2008), 
the tidal influence of
the dark halo of the main galaxy in a group reduces the central surface brightness of a
satellite and shortens its size.  Torrealba et al. [27] %Torrealba et al. (2019) 
noted that
N-body modeling a strong tidal stripping of a diffuse companion can explain the observed
properties of Antlia\,2-type galaxies.

More in-depth observations of the mentioned triplets of galaxies both in the optical range and
in the neutral hydrogen line could clarify the problem of the supposed
connection between the cold kinematics of the group’s satellites and the presence of LDD
galaxies in it.

{\bf Acknowledgements.} This work has made use of the DESI Legacy Imaging Surveys, the Lyon
Extragalactic Database and the revised version of the Local Volume database.
IDK and AEN are supported by the Russian Science Foundation grant 24--12--00277.

REFERENCES

1. I.D.Karachentsev, D.I.Makarov, E.I.Kaisina, Astron. J., {\bf 145}, 101, 2013.

2. P.van Dokkum, R.Abraham, A.Merritt et. al., Astropys. J.,  {\bf 798L}, 45, 2015.

3. J.C.Mihos, P.Harding, J.J.Feldmeier et al., Astropys. J., {\bf 834}, 16, 2017.

4. R.P.Munoz, P.Eigenthaler, T.H.Puzia  et al.,  Astropys. J., {\bf 813}, L15, 2015.

5. J.Koda, M.Yagi, H.Yamanoi, Y.Komiyama,  Astropys. J., {\bf 807}, L2, 2015.

6. D.Crnojevic, D.J.Sand, K.Spekkens et al.,  Astropys. J., {\bf 823}, 19, 2016.

7. E.Toloba, D.J.Sand, K.Spekkens et al., Astropys. J., {\bf 816}, L5, 2016.

9. A.Merritt, P.van Dokkum, S.Danieli et al., Astropys. J., {\bf 833}, 168, 2016.

9. D.Zaritsky, R.Donnerstein, A.Dey et al., Astrophys. J. Suppl., {\bf 267}, 27, 2023.

10. J.Roman, I.Trujillo, Mon. Not. Roy. Astron. Soc., {\bf 468}, 4039, 2017.

11. I.D.Karachentsev, L.N.Makarova, M.E.Sharina,  V.E.Karachentseva, 
 Astrophysical Bulletin, {\bf 72}, 376, 2017.

12. A.Dey, D.J.Schlegel, D.Lang et al., Astron. J., {\bf 157}, 168, 2019.

13. D.Makarov,  I.Karachentsev,  Mon. Not. Roy. Astron. Soc., {\bf 412}, 2498, 2011.

14. I.Karachentsev, O.Kashibadze, Astronomische Nachrichten, {\bf 342}, 999, 2021.

15. V.E.Karachentseva, I.D.Karachentsev, M.E.Sharina, Astrophysics,
{\bf 53}, 462, 2010.

16. M.E.Sharina,  I.D.Karachentsev, V.E.Karachentseva,  in The Zeldovich Universe: Genesis and Growth of the Cosmic Web, ed. R. van de
Weygaert, S. Shandarin, E. Saar, \& J. Einasto, {\bf 308}, 473, 2016.

17. D.I.Makarov,  I.D.Karachentsev,  Astrophysical Bulletin, {\bf 64}, 24, 2009.

18. D.Makarov,  P.Prugniel, N.Terekhova,  H.Courtois, I.Vauglin,  Astron. Astrophys.,
{\bf 570}, A13, 2014.

19. E.J.Shaya,  R.B.Tully, Y.Hoffman,  D.Pomar\'{e}de , Astropys. J., {\bf 850}, 207, 2017.

20. J.L.Tonry,  A.Dressler,  J.P.Blakeslee et al.,  Astropys. J., {\bf 546}, 681, 2001.

21. T.M.C.Abbott, M.Adam\'{o}w, M.Aguena et al., 2021, Astropys. J. Suppl., {\bf 255}, 20, 2021.

22. N.C.Amorisco,  A.Loeb,   Mon. Not. Roy. Astron. Soc., {\bf 459}, L51, 2016.

23. A.Di Cintio, C.B.Brook, A.Dutton  et al., Mon. Not. Roy. Astron. Soc., {\bf 466}, L1, 2017.

24. P.van Dokkum, R.Abraham,  J.Brodie et al.,  Astropys. J., {\bf 828}, L6, 2016.

25. I.D.Karachentsev, E.I.Kaisina,  Astrophysical Bulletin, {\bf 77}, 372, 2022.

26. S.Okamoto,  A.M.N.Ferguson, N.Arimoto et al., Astropys. J., {\bf 967L}, 240, 2024.

27. G.Torrealba, V.Belokurov,  S.E.Koposov et al.,  Mon. Not. Roy. Astron. Soc., {\bf 488}, 2743, 2019.

28. N.F.Martin,  R.A.Ibata, G.F.Lewis et al., Astropys. J., {\bf 833}, 167, 2016.

29. J.Pe\~{n}arrubia,  J.F.Navarro, J. A.W.McConnachie,   Astropys. J., {\bf 673}, 226, 2008.

   \end{document}